# Simplified Self-homodyne Coherent System Based on Alamouti Coding and Digital Subcarrier Multiplexing

Wei Wang, Dongdong Zou, Zhenpeng Wu, Qi Sui, Xingwen Yi, Fan Li, Chao Lu, and Zhaohui Li

*Abstract*—Coherent technology inherent with more available degrees of freedom is deemed a competitive solution for next-generation ultra-high-speed short-reach optical interconnects. However, the fatal barriers to implementing the conventional coherent system in short-reach optical interconnect are the cost, footprint, and power consumption. Self-homodyne coherent system exhibits its potential to reduce the power consumption of the receiver-side digital signal processing (Rx-DSP) by delivering the local oscillator (LO) from the transmitter. However, an automatic polarization controller (APC) is inevitable in the remote LO link to avoid polarization fading, resulting in additional costs. To address the polarization fading issue, a simplified self-homodyne coherent system is proposed enabled by Alamouti coding in this paper. Benefiting from the Alamouti coding between two polarizations, a polarization-insensitive receiver only including a 3dB coupler, a 90° Hybrid, and two balanced photodiodes (BPDs) is sufficient for reception. Meanwhile, the APC in the LO link is needless, simplifying the receiver structure significantly. Besides, the digital subcarrier multiplexing (DSCM) technique is also adopted to relax the computational complexity of the chromatic dispersion compensation (CDC), which is one of the dominant power consumption modules in Rx-DSP. The transmission performance of 50Gbaud 4-subcarrier 16/32QAM (4SC-16/32QAM) DSCM signal based on the proposed simplified self-homodyne coherent system is investigated experimentally. The results show that the bit-error-ratio (BER) performance degradation caused by CD can be solved by increasing 4 taps in the equalizer for 80km single mode fiber (SMF) transmission without individual CDC, which operates in a low-complexity manner.

## I. Introduction

To support the development of the digital economy, the demand for transmission capacity of data center interconnects (DCIs) increases rapidly. For short-reach optical interconnect applications, the cost and power consumption are the main constraints [1, 2]. Nowadays, the intensity modulation and direct detection (IM/DD) system is still the mainstream for short-reach optical interconnects due to the merits of low cost, low power consumption, and simple system architecture [3-5]. However, the next-generation short-reach Ethernet links are aiming at 800GbE or even 1.6TbE [6, 7]. Meeting such high data rate requirements is a considerable challenge for IM/DD due to its limited spectral efficiency and poor reception linearity [8]. In contrast, coherent reception seems to be an attractive candidate, which has the ability to achieve larger capacity benefiting from its high spectral efficiency, high linearity, and high receiver sensitivity [9-11]. However, the cost, footprint, and power consumption of the conventional coherent transceivers are the fatal barriers to substituting IM/DD in short-reach optical communication scenarios [12]. To promote the introduction of coherent technology into short-reach optical interconnects, the redesign of conventional coherent systems is inevitable, and research on the simplification of coherent technology has received widespread attention in recent years [13-28].

Generally, the simplification of coherent technology can be considered in two folds, namely the complexity of digital signal processing (DSP) algorithms and system architecture. In the conventional coherent system, a local oscillator (LO) is utilized at the receiver side (Rx-side) for signal detection, and the frequency and phase of the optical carrier emitting from LO are not strictly locked with that of the transmitter side (Tx-side) laser. Therefore, stringent requirements of wavelength stability and narrow linewidth of the transceiver lasers are indispensable, and temperature control is also needed to avoid wavelength shifts. Subsequently, the system cost and power consumption increase. Since the frequency offset and phase noise are unavoidable in the conventional coherent system, the frequency offset estimation and compensation based on fast Fourier transform (FFT) [29] and phase noise compensation based on blind phase search (BPS) [30] are required in the receiver side

Manuscript received XXX XXXX; revised XXX, XXXX; accepted XXX, XXXX. This work is partly supported by the National Key R&D Program of China (2018YFB1800902); National Natural Science Foundation of China (U2001601, 62271517, 62035018); Guangdong Basic and Applied Basic Research Foundation (2023B1515020003), Open Fund of IPOC (BUPT) (IPOC2020A010), Fundamental and Applied Basic Research Project of Guangzhou City (202002030326), Local Innovation and Research Teams Project of Guangdong Pearl River Talents Program (2017BT01X121). (*Corresponding Author: Fan Li*)

W. Wang, D. Zou, Z. Wu, X. Yi, F. Li, C. Lu and Z. Li are with are with School of Electronics and Information Technology, the Guangdong Provincial Key Laboratory of Optoelectronic Information Processing Chips and Systems, Sun Yat-Sen University, Guangzhou 510275, China and Southern Marine Science and Engineering Guangdong Laboratory (Zhuhai), Zhuhai 519000, China (e-mail: lifan39@mail.sysu.edu.cn).

Qi Sui is with the Southern Marine Science and Engineering Guangdong Laboratory, Zhuhai 519000, China (e-mail: suiqi@sml-zhuhai.cn).

DSP, which incurs a large amount of DSP overhead. To address this issue, several schemes are proposed to avoid or reduce the complexity of frequency offset and phase noise compensation. Analog coherent detection is proposed to achieve frequency offset and phase noise corrections in the analog domain. An optical phase-locked loop (OPLL) circuitry is required to lock the frequency and phase of the LO to the incoming signal [13, 14]. In addition to using OPLL to remove the frequency and time-varying phase offset between the transmitted signal and LO, in Ref. [15], an analog circuit for carrier phase recovery and compensation is implemented after detection. Recently, self-coherent systems have attracted much attention, in which the LO and signal come from the same laser. A stimulated Brillouin scattering (SBS) based optical carrier recovery method is proposed to achieve self-coherent reception [16-19]. In this scheme, the LO is regenerated from the incoming signal itself using nonlinear SBS. However, an SBS-based optical carrier recovery loop is required, which is mainly composed of a gain medium, a circulator, a frequency shifter, and an amplifier. Besides, the amplified carrier frequency is approximately 11GHz away from the pump optical frequency. In this circumstance, the Rx-side laser is saved at the cost of an extra optical carrier recovery loop.

Although the abovementioned schemes can reduce the computational complexity of Rx-DSP, the receiver structure has been complicated to some extent to obtain a locked LO. Recently, the self-homodyne coherent scheme realized by remotely delivering an LO from the same Tx-side laser has attracted much attention [20, 21]. There is no frequency offset between LO and the transmitted signal, and phase noise is introduced when the length of two fiber links is mismatched [21]. However, due to the random polarization variation of the remote LO, the received signal will suffer from polarization fading. Generally, a complicated automatic polarization controller (APC) is indispensable for the remote LO to track its polarization state [22], introducing extra cost. To avoid the use of APC in a self-homodyne coherent system, a hybrid polarization-diversity coherent receiver (composed of a hybrid single-polarization coherent receiver and a Stokes vector receiver) and an integrated complementary polarization-diversity coherent receiver including three 90° Hybrids and six balanced photodiodes (BPDs) are proposed in Ref. [23] and Ref. [24], respectively. Obviously, the receiver structure becomes more complicated while avoiding optical polarization control.

To remove the APC in the LO link while avoiding increasing the complexity of the receiver structure, a simplified self-homodyne coherent system enabled by Alamouti polarization-time block coding is proposed in this paper. According to previous works [26-28], the conventional coherent receiver can be changed into a polarization-insensitive by adopting Alamouti coding. In Ref. [26], a simple coherent receiver only including a 3dB coupler and a BPD without any polarization-dependent devices is achieved by utilizing Alamouti coding and heterodyne detection technique. Inspired by the polarization-insensitive property brought by Alamouti coding, we propose to solve the polarization fading issue in the self-homodyne coherent system without complicated APC by utilizing Alamouti coding. Meanwhile, the receiver is also simplified compared to the conventional coherent receiver, which only consists of a 3dB coupler, a 90° Hybrids, and two BPDs. Except for the simplification in the receiver structure of the self-homodyne coherent system, reducing the computational complexity of the Rx-DSP is another objective of this work. Although the power-hungry modules as frequency offset and phase noise compensation can be removed or simplified due to the inherent merit of the self-homodyne coherent system, chromatic dispersion compensation (CDC) is also a significant power consumption portion in Rx-DSP. It is well known that the length of the digital filter required for CDC is proportional to the square of the baud rate [31]. Therefore, digital subcarrier multiplexing (DSCM) technology is utilized to simplify the CDC procedure in the proposed system. Since the key idea of DSCM technology is to divide a high baud rate single carrier signal into several low baud rate subcarriers without spectrum overlap [31-34], the compensation complexity for each subcarrier is much lower than that for the entire single carrier signal. In this paper, the effectiveness of the proposed simplified self-homodyne coherent scheme is verified experimentally. The bit error rate (BER) performance of 50Gbaud 4-subcarrier 16-ary quadrature amplitude modulation (4SC-16QAM), and 50Gbaud 4SC-32QAM DSCM signals in optical back-to-back (OBTB) and 80km single-mode fiber (SMF) transmission is investigated. The experimental results show that the BER can be below the hard-decision forward error correction (HD-FEC) threshold with optical signal-to-noise ratios (OSNR) of 25dB and 34dB for 50Gbaud 4SC-16QAM and 50Gbaud 4SC-32QAM without complicate frequency offset and phase noise compensation in OBTB transmission. Besides, according to the experimental results, the BER degradation due to CD can be settled by increasing 4 taps in the subsequent equalizer for 80km SMF transmission without an individual CDC module. The computational complexity of dispersion compensation is also discussed in detail. The results show that the proposed method utilizing the DSCM signal and addressing CD in the equalizer can achieve significant complexity reduction, which further reduces the complexity of Rx-DSP.

The rest of the paper is organized as follows. Section II gives the operation principles of the proposed method and verifies its polarization-insensitive reception by simulation. Section III shows the experimental setup as well as the results discussion. In section IV, the complexity analysis of dispersion compensation is discussed. Finally, the conclusion is drawn in section V.

## II. Theory

### A. Polarization insensitive reception enabled by Alamouti coding

Different from the conventional dual-polarization (DP) modulation scheme, the key idea of Alamouti polarization-time block coding is to send the same information during two symbol durations, which is interleaved between two polarizations. Fig. 1(a) shows the conventional DP modulation, in which the signals carried on X- and Y-polarization are irrelevant, while Fig. 1(b) shows the received signal structure. However, the polarization state of the remote LO rotates arbitrarily after fiber transmission as shown in Fig. 1(c), and the received signal will suffer from polarization fading. This problem can be solved by utilizing Alamouti coding, and its operation principle is shown in Fig. 1(d). The signals on two polarizations are denoted as $E_x = \{s_1, -s_2^*, s_3, -s_4^*, \cdots\}$ and $E_y = \{s_2, s_1^*, s_4, s_3^*, \cdots\}$, where * represents the conjugate operation. Considering the channel transmission matrix $\mathbf{H} = [h_{xx}, h_{xy}; h_{yx}, h_{yy}]$ and phase noise $\theta$ induced by fiber length mismatch, the received signal of one polarization in Fig. 1(e), for example, X-polarization can be expressed as [28]:

$$\begin{aligned} s_{x1}^{'} &= h_{xx} s_1 e^{j\theta} + h_{xy} s_2 e^{j\theta} \\ s_{x2}^{'} &= -h_{xx} s_2^* e^{j\theta} + h_{xy} s_1^* e^{j\theta} \end{aligned} \quad (1)$$

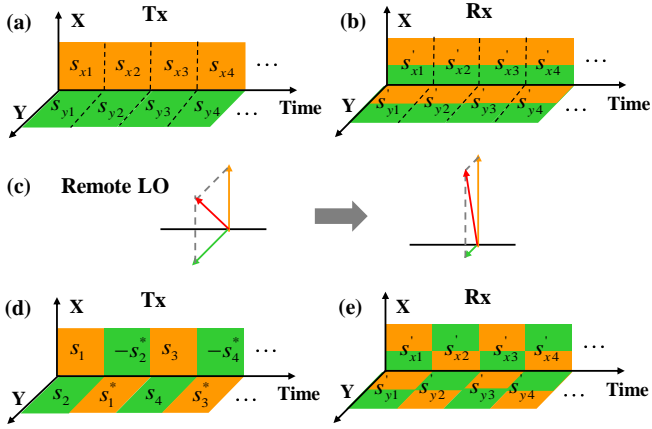

Fig. 1. Conventional DP modulation (a) Tx-side, (b) Rx-side. (c) Polarization rotation of the remote LO. Alamouti coding scheme (d) Tx-side, (e) Rx-side.

As shown in equation (1), polarization crosstalk exists in the received signal. The signal recovery can be achieved by:

$$\begin{bmatrix} s_1 \\ s_2 \end{bmatrix} = \begin{bmatrix} h_{xx} e^{j\theta} & h_{xy} e^{j\theta} \\ h_{xy}^* e^{-j\theta} & -h_{xx}^* e^{-j\theta} \end{bmatrix}^{-1} \begin{bmatrix} s_{x1}^{'} \\ s_{x2}^{'*} \end{bmatrix} \quad (2)$$

The equalizer structure is shown in Fig. 2. And $p$ represents a single-tap phase factor to compensate for the phase noise. The equalization process can be expressed as:

$$\begin{aligned} \mathbf{s}_{xo}(n) &= \mathbf{w}_{11} \mathbf{s}_{xo}^{'} p + \mathbf{w}_{12} \mathbf{s}_{xe}^{'*} p^* \\ \mathbf{s}_{xe}(n) &= \mathbf{w}_{21} \mathbf{s}_{xo}^{'} p + \mathbf{w}_{22} \mathbf{s}_{xe}^{'*} p^* \end{aligned} \quad (3)$$

where $\mathbf{s}_{xo}^{'}$ and $\mathbf{s}_{xe}^{'}$ are the odd and even sequences of the received signal, and $\mathbf{w}_{11}$, $\mathbf{w}_{12}$, $\mathbf{w}_{21}$, $\mathbf{w}_{22}$ are the finite impulse response (FIR) filters to address polarization crosstalk and realize channel equalization. Both the tap coefficients of the FIR filters and the phase factor are updated by the least mean square (LMS) criterion. The update of tap coefficients of FIR filters can be expressed as:

$$\begin{aligned} \mathbf{w}_{11/21} &= \mathbf{w}_{11/21} + \frac{\mu |p|}{p} e_{o/e} \mathbf{s}_{xo}^{'*} \\ \mathbf{w}_{12/22} &= \mathbf{w}_{12/22} + \frac{\mu |p|}{p^*} e_{o/e} \mathbf{s}_{xe}^{'} \end{aligned} \quad (4)$$

where $\mu$ is the step size to control the convergence speed. And $e_{o/e}$ is the error signal, which is obtained by:

$$e_{o/e} = d_{o/e} - s_{xo/xe} \quad (5)$$

where $d_{o/e}$ is the training symbol or decided ideal symbol. As for the phase factor, its update rule is:

$$\begin{aligned} p_1 &= p_1 + \mu_p e_o \left(\mathbf{w}_{11} \mathbf{s}_{xo}^{'}\right)^* \\ p_2 &= p_2 + \mu_p e_o \left(\mathbf{w}_{12} \mathbf{s}_{xe}^{'*}\right)^* \\ p &= (p_1 + p_2^*)/2 \end{aligned} \quad (6)$$

where $\mu_p$ is the step size to control the convergence speed.

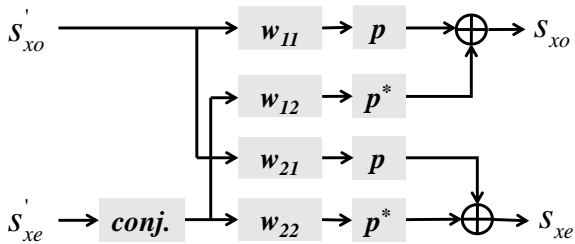

Fig. 2. Equalizer structure.

According to the above description, the transmitted signal can be recovered by detecting in any polarization state, indicating that a polarization-insensitive receiver including only a 3dB coupler, a 90° Hybrid, and two BPDs are sufficient. Therefore, the polarization fading issue in the conventional self-homodyne coherent system can be avoided without APC.

### B. Digital subcarrier multiplexing

As discussed in Ref. [31], the length of impulse response expanded due to dispersion is quadratically related to the signal baud rate. Thus, the signal with a low baud rate will be less affected by dispersion. From this perspective, utilizing the DSCM technique has the merit of simplifying the CDC procedure, since the key idea of DSCM is to divide a high baud rate signal into several low baud rate subcarrier signals. Fig. 3 gives the spectrum schematic diagram of the single-carrier signal and DSCM signal with 4 subcarriers. For the convenience of later description, the subcarriers located at low frequency are labeled as subcarrier 1 (SC1) and subcarrier 2 (SC2), while the subcarriers located at high frequency are labeled as subcarrier 3 (SC3) and subcarrier 4 (SC4) respectively. As shown in Fig. 3, the overall spectral occupancy of these two signals is the same with the same roll-off factor since the gaud band is unnecessary for the DSCM signal. Thus, the DSCM technology can guarantee spectral efficiency while it is more robust to CD. And the CDC is done separately for each subcarrier in a low-complexity manner. In section III, the robustness of single-carrier signal and DSCM signal to CD will be further investigated and discussed by experiment.

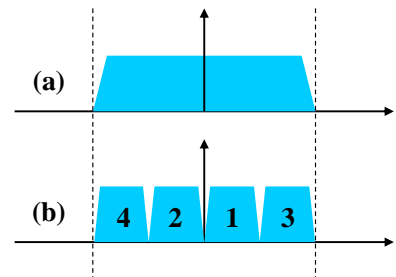

Fig. 3. Spectrum schematic diagram of (a) single-subcarrier signal and (b) 4-subcarrier DSCM signal.

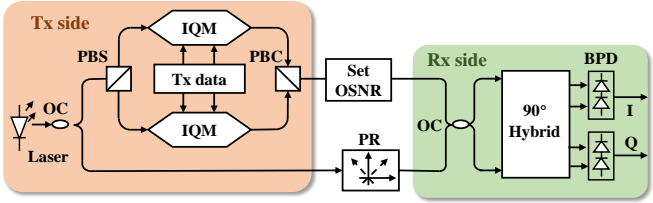

Fig. 4. Simulation setup.

*C. Simulation analysis*

In this subsection, the polarization-insensitive characteristic of the proposed simplified self-homodyne coherent system is validated by simulation. The simulation is carried out by MATLAB and VPItransmissionMaker. Fig. 4 gives the simulation setup of the proposed simplified self-homodyne coherent system. At the transmitter side, the optical carrier with a linewidth of 100kHz is divided into two tributaries. One is used for signal modulation, and the other is served as LO. To achieve DP modulation, one branch of the optical carrier is split into two orthogonal polarizations by PBS and then injected into two IQ modulators (IQM) for signal modulation. After that, a polarization beam combiner (PBC) is used for the combination of two orthogonal signals. In the signal transmission link, an OSNR setting module is implemented to adjust the OSNR of the system, and the OSNR is set as 26dB in this simulation. To verify the polarization-insensitive performance of the proposed system, a polarization rotator (PR) is added in the remote LO link to change the polarization state of the LO. At the receiver side, the coherent receiver consists of a 2×2 coupler, a 90° Hybrid, and two BPDs. The simulation investigates the BER performance of 50Gbaud 4SC-16QAM DSCM signal in OBTB transmission by varying rotation angles both in azimuth angle and elevation angle of LO. The simulation results are shown in Fig. 5, and both the azimuth angle and elevation angle are swept from -90° to 90°. As shown in Fig. 5, the $Q^2$ factor varies within a small range regardless of any polarization rotation, reflecting the robustness of the changes in polarization states. The simulation results imply that the proposed simplified self-homodyne coherent system enabled by Alamouti coding can effectively avoid power fading in the absence of APC.

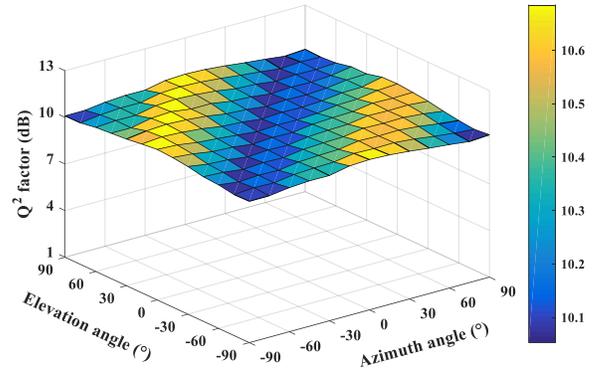

Fig. 5. Simulation analysis of the influence of polarization rotation.

### III. EXPERIMENTAL SETUP AND RESULTS

*A. Experimental setup*

The experimental setup of the proposed simplified self-homodyne coherent system and the corresponding DSP flow are given in Fig. 6. At the transmitter side, the optical carrier emitting from a laser with optical power and wavelength of 16dBm and 1550nm is divided into two branches by a polarization-maintaining optical coupler with a splitting ratio of 80:20. One branch with 80% optical power is injected into a DP-IQM for optical signal modulation. While the other branch with 20% optical power of the optical carrier is delivered to the receiver side to serve as an LO. An Erbium-doped fiber amplifier (EDFA) is used to boost the LO after transmission, and the optical power of LO after amplification is 16dBm. The DSCM signal is generated offline in MATLAB, and the Tx-DSP is shown in Fig. 6. The pseudo-random binary sequence (PRBS) with length of $2^{20}$ is used and mapped into two sets of QAM symbols for X- and Y-polarization respectively. In this paper, the performance of the 4SC DSCM signal is investigated, thus each polarization contains 4 subcarriers. Then, Alamouti encoding is executed between two polarizations for each subcarrier, and the coded signals are Nyquist shaped by a root-raised-cosine (RRC) filter with a roll-off factor of 0.1. Afterward, the 4 subcarriers are multiplexed to obtain the DSCM signal. To ensure the spectral efficiency of the signal, there is no frequency interval between adjacent subcarriers in our experiment. The signal is then resampled to 64GSa/s to match the sampling rate of the Keysight M8195 arbitrary waveform generator (AWG). The outputs of AWG are then amplified and loaded into the DP-IQM. And the optical DSCM signal is boosted by an EDFA before being injected into the fiber.

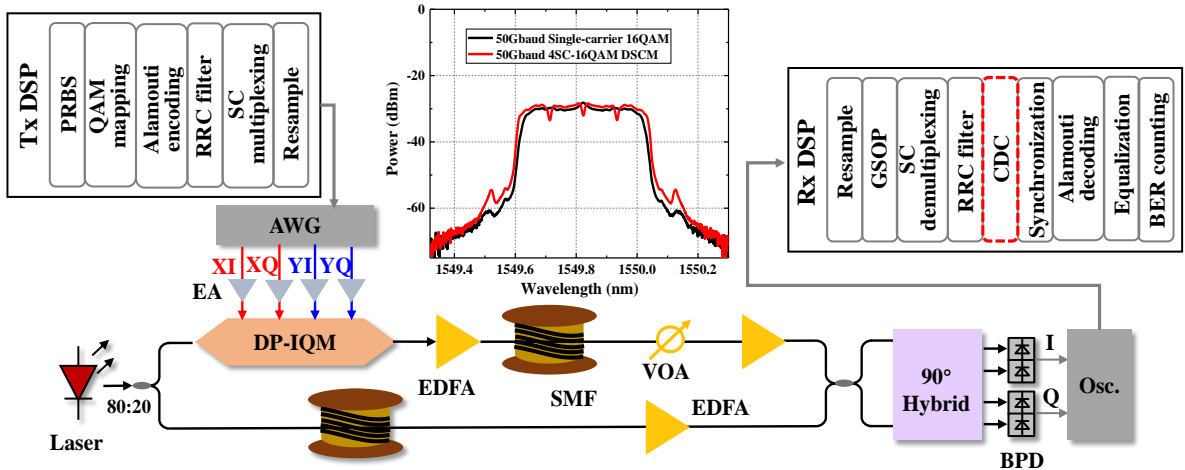

Fig. 6. Experimental setup of the proposed simplified self-homodyne coherent system.

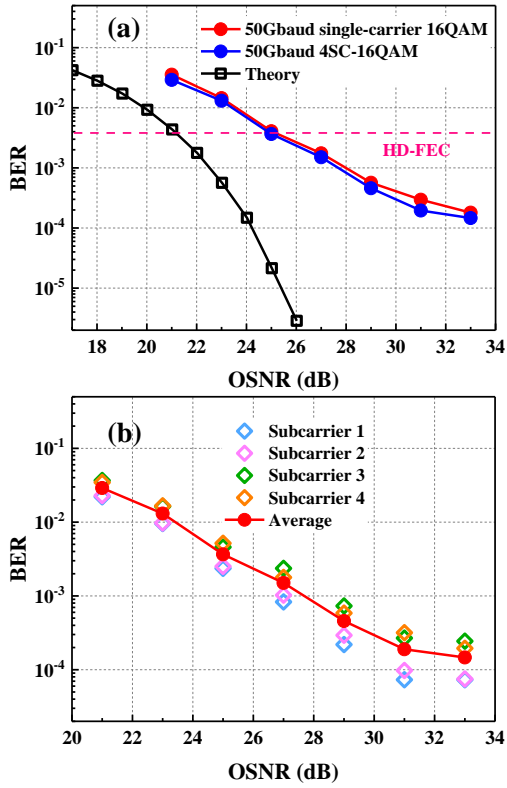

Fig. 7. BER versus OSNR in OBTB transmission. (a) Comparison between 50Gbaud 4SC-16QAM and single-carrier 16QAM, (b) BER of each subcarrier.

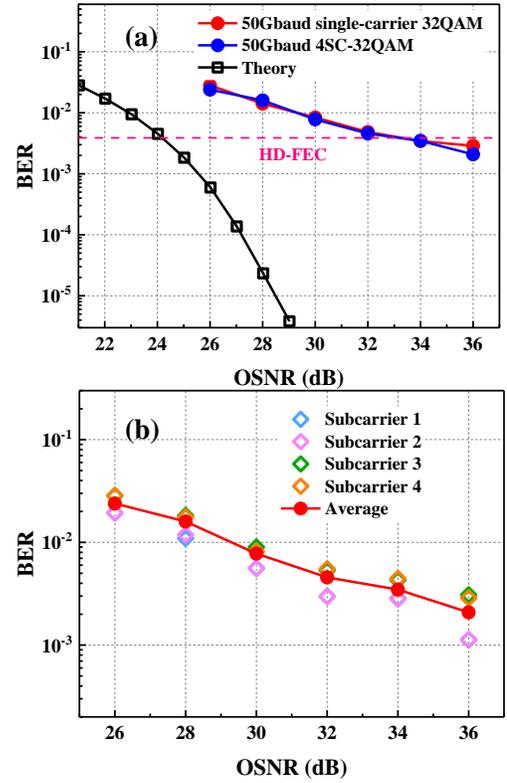

Fig. 8. BER versus OSNR in OBTB transmission. (a) Comparison between 50Gbaud 4SC-32QAM and single-carrier 32QAM, (b) BER of each subcarrier.

In this experiment, the performance of 80km SMF transmission is investigated. A variable optical attenuator (VOA) and an EDFA are used to introduce amplified spontaneous emission (ASE) noise and adjust the OSNR. Due to the adoption of Alamouti coding, only a 3-dB coupler, a 90° Hybrid, and two BPDs are utilized to achieve polarization-insensitive reception. Subsequently, the APC is also avoidable for the remote LO regardless of variation in polarization state. Thus, the receiver complexity is reduced significantly compared to the conventional coherent receiver. The received optical power is set to be -13dBm, and the signal detected by BPDs is captured by a Lecory oscilloscope (Osc) operating at 80GSa/s with a cut-off bandwidth of 36GHz. Then the captured signal is processed in offline DSP. Since the LO comes from the same laser in this self-homodyne coherent system, the frequency offset estimation and compensation can be saved in the Rx-DSP, and the phase noise induced by fiber mismatch is addressed by one-tap phase factor in the equalizer, which effectively reduces the complexity of the Rx-DSP to a certain extent. In the Rx-DSP, the signal is resampled to 2 samples per symbol, and the Gram-Schmidt orthogonalization procedure (GSOP) is implemented. Then, subcarriers are de-multiplexed and each subcarrier is processed separately in the following procedure including RRC filter, synchronization, Alamouti decoding, and equalization. Finally, the BER is counted.

*B. Transmission performance analysis*

At first, the BER performance of Alamouti coded DSCM signal based on the proposed simplified self-homodyne coherent system is investigated in OBTB transmission. Fig. 7(a) shows the average BER performance of the 50Gbaud 4SC-16QAM DSCM signal under different OSNRs, which is compared to the 50Gbaud single-carrier 16QAM signal. As shown in Fig. 7(a), the BER curves of the DSCM signal and single-carrier signal almost overlap. The required OSNR is about 25dB at the HD-FEC threshold of 3.8e-3. The theory BER curve is also plotted for comparison, and about 4dB OSNR penalty can be observed at the HD-FEC threshold. It can be attributed to electrical noise (induced by electric amplifier, PD), nonlinear impairment of devices, and other unbalanced impairments in the practical system. The optical spectrums of 50Gbaud 4SC-16QAM and single-carrier 16QAM signals are given in Fig. 6. Fig. 7(b) shows the BER performance of each subcarrier. Due to the system bandwidth limitation, the subcarriers located at high frequency will suffer more serious damage. Thus, the BER performance of SC3 and SC4 is worse than that of SC1 and SC2. While the BER performance of SC1(3) and SC2(4) is almost the same. Fig. 8 gives the BER curves of the 50Gbaud 4SC-32QAM DSCM signal. According to the experimental results, the BER of the 50Gbaud 4SC-32QAM DSCM signal is almost the same as that of the 50Gbaud single-carrier 32QAM signal, and the required OSNR is about 34dB at the HD-FEC threshold. Compared to the case of 16QAM, the OSNR penalty of 50Gbaud 32QAM is much larger, since the signal with higher modulation format is more sensitive to system impairments. The BER of each subcarrier of 50Gbaud 4SC-32QAM DSCM signal is also given in Fig. 8(b). Similar to the case of 50Gbaud 4SC-16QAM, the BER performance of the subcarrier located at high frequency is worse than that of the subcarrier located at low frequency.

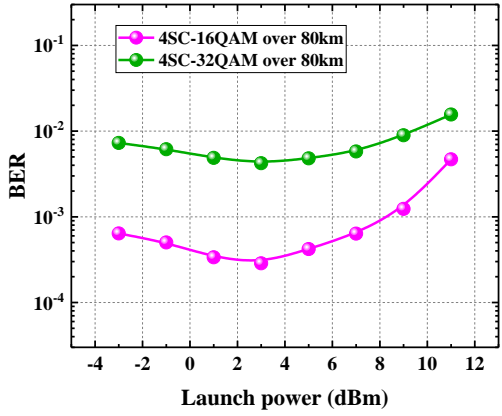

Fig. 9. BER versus launch power over 80km SMF transmission.

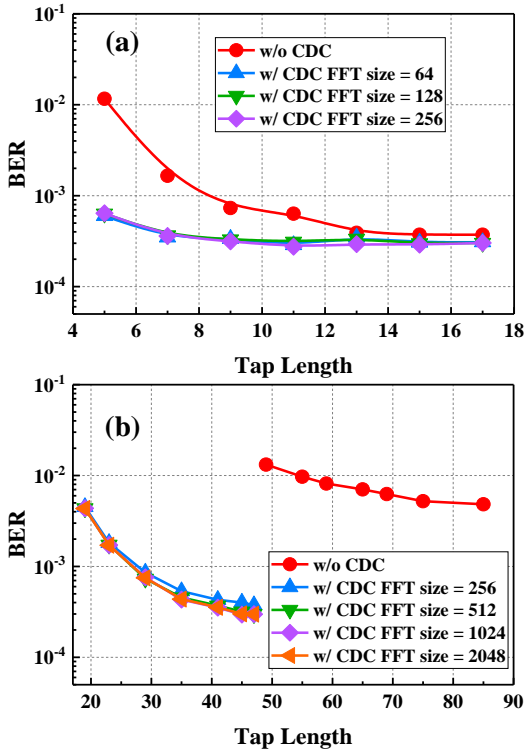

Fig. 10. BER versus tap length over 80km SMF transmission (a) 50Gbaud 4SC-16QAM DSCM signal, (b) 50Gbaud single-carrier 16QAM signal.

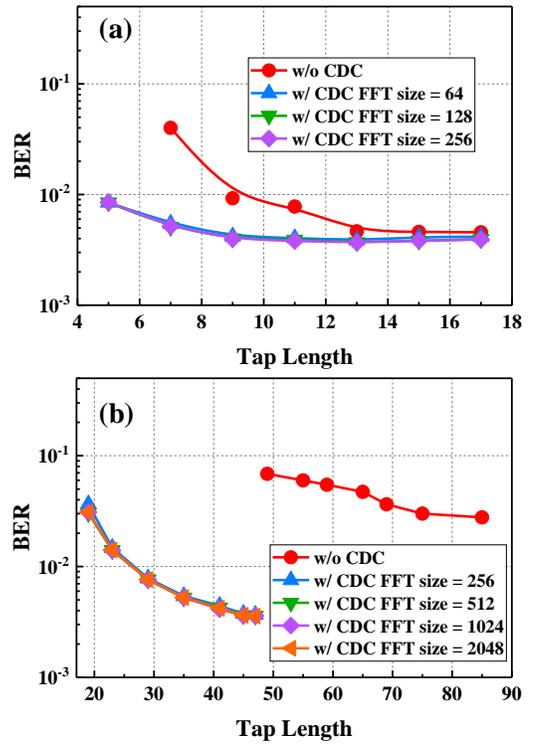

Fig. 11. BER versus tap length over 80km SMF transmission (a) 50Gbaud 4SC-32QAM DSCM signal, (b) 50Gbaud single-carrier 32QAM signal.

Then, the BER performance of 50Gbaud 4SC-16QAM and 50Gbaud 4SC-32QAM DSCM signals over 80km SMF is also tested. Fig. 9 gives the measured BER versus launch power over 80km SMF. As shown in Fig. 9, the optimal launch power of 50Gbaud 4SC-16QAM and 4SC-32QAM DSCM signals in 80km SMF transmission is about 3dBm. Then, the CD tolerance of the DSCM signal is discussed. Fig. 10(a) compares the BER performance of 50Gbaud 4SC-16QAM DSCM signal with and without frequency domain CDC (FD-CDC) by varying the tap length of the equalizer. The FD-CDC is achieved by the overlap-and-save method block-to-block using FFT and inverse FFT (IFFT), and different FFT sizes are considered in our experiment. As shown in Fig. 9(a), the BER performance with FFT sizes of 64, 128, and 256 is almost the same, and approximately 9 taps are required for 50Gbaud 4SC-16QAM DSCM signal to achieve the best performance over 80km SMF transmission when FD-CDC is implemented. In the absence of individual CDC, about 13 taps are required. It indicates that the procedure of CDC realized in the frequency domain can be saved out only increases 4-tap longer of the following equalizer for 80km SMF transmission, operating in a lower complexity manner. For comparison, the BER performance of 50Gbaud single-carrier 16QAM signal versus tap length with and without FD-CDC is given in Fig. 10(b). According to the experimental results, the required tap length is about 45 when FD-CDC is implemented for the 50Gbaud single-carrier 16QAM signal over 80km SMF transmission. However, as for the signal without CDC, increasing the tap length of the equalizer is also unable to diminish the penalty induced by CD due to the non-optimal convergent tap coefficients for dispersion. The BER performance of 50Gbaud 4SC-32QAM DSCM and single-carrier 32QAM signals versus tap length over 80km SMF with and without CDC is given in Fig. 11. As shown in Fig. 11(a), the required tap length of 50Gbaud 4SC-32QAM DSCM signal is 9 with FD-CDC. As for the case without individual CDC, the required tap length increases to 13 to eliminate the performance penalty. In addition, the BER performance of the 50Gbaud single-carrier 32QAM signal with different tap length is also investigated, and the results are given in Fig. 11(b). Same as the case of 50Gbaud single-carrier 16QAM signal, about 45 taps are required to achieve the best performance utilizing FD-CDC, while significant performance penalty can be observed though large tap length is used. According to the experimental results, the penalty caused by CD can be addressed by increasing several taps in the equalizer for the 4SC-16/32QAM DSCM signal, which operates in a lower complexity manner compared to FD-CDC. A detailed complexity comparison is given in section IV. Thus, the proposed simplified self-homodyne coherent method provides a characteristic of low complexity not only in the structure but also in DSP flow.

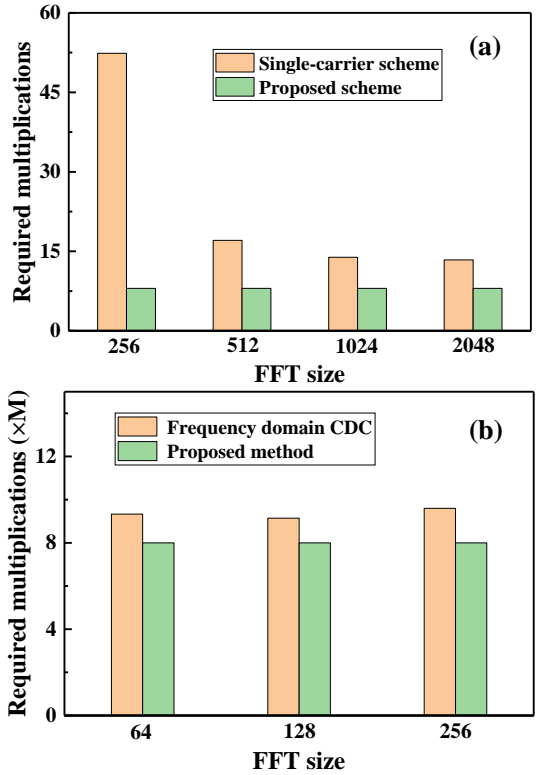

Fig. 12. Computational complexity comparison between (a) single-carrier scheme and proposed scheme, (b) DSCM signal using frequency domain CDC and proposed scheme.

## IV. COMPLEXITY ANALYSIS

In this section, the required computational complexity for dispersion compensation of single-carrier signal with FD-CDC, DSCM signal with FD-CDC, and DSCM signal with the proposed method is discussed. Assume the length of the processed data is $M$. For frequency domain CDC achieved by the overlap-and-save method, the computational complexity of FFT or IFFT for each block is $(N\log_2 N)/2$ multiplications and $N\log_2 N$ additions with an FFT size of $N$. Besides, $N$ multiplications are required to multiply by the corresponding phase factor in the frequency domain for dispersion compensation. Thus, the required multiplications and additions are $N(1+\log_2 N)$ and $2N\log_2 N$ for each block. With data length of $M$, the block number is $M/(N-2N_{OL})$, where $N_{OL}$ is the overlap length. Then, the total computational complexity of the FD-CDC is $MN(1+\log_2 N)/(N-2N_{OL})$ multiplications and $2MN\log_2 N/(N-2N_{OL})$ additions. In our experiment, the overlap length $N_{OL}$ is 106 for the single-carrier signal after 80km SMF transmission, while the overlap length adopted for the DSCM signal is 8. As for our proposed simplified method utilizing the DSCM signal, the performance penalty induced by CD is addressed in the equalizer by increasing 4 taps for 80km transmission without individual CDC. Thus, the additional multipliers and adders added to the equalizer are both 16. According to the equalizer structure given in section II-A, the received data sequence is divided into odd and even sequences as the input of the 2×2 equalizer. Then the required computational complexity to address dispersion in the proposed method is $8M$ multiplications and $8M$ additions. To provide an intuitive comparison, Figs. 12(a) and 12(b) give the specific computational complexity of per symbol in dispersion compensation for single-carrier with FD-CDC and DSCM signal with FD-CDC under different FFT sizes, which also compare to the proposed method. As shown in Fig. 12(a), significant computational complexity reduction can be achieved for the proposed method compared to the single-carrier scheme. According to Fig. 12(b), the required multiplications for the proposed method to deal with dispersion are fewer than the DSCM signal utilizing FD-CDC.

## V. CONCLUSION

In this work, a simplified self-homodyne coherent system is proposed to achieve the simplification both in system architecture and complexity of Rx-DSP. Alamouti coding is adopted to deal with the power fading issue due to the arbitrary polarization state variation of the remote LO in the self-homodyne coherent system so that the APC can be removed. Meanwhile, the receiver can also be simplified to a polarization-insensitive receiver consisting of a 3dB coupler, a 90° Hybrid, and two BPDs, which reduces the complexity of the receiver significantly. The transmission performance of 50Gbaud 4SC-16QAM and 50Gbaud 4SC-32QAM DSCM signals are experimentally investigated. Since the LO comes from the same laser at the transmitter side, the frequency offset compensation is omitted and phase noise compensation is realized by adding a one-tap phase term in the Rx-side equalizer. Besides, DSCM technology is used to further reduce the CDC procedure. The experimental results show that the performance penalty caused by CD can be addressed by increasing 4 taps in the subsequent equalizer for both 50Gbaud 4SC-16QAM and 4SC-32QAM DSCM signals over 80km SMF transmission. As a result, the proposed simplified self-homodyne coherent system simplifies both system structure and Rx-DSP, providing the possibility for the application of coherent technology in the next-generation short-reach optical interconnects.